\documentstyle[12pt,psfig]{article}
\begin{document}

\hfill PUPT-1707

\hfill hep-th/9705210

\vspace{1.5in}

\begin{center}

{\large\bf Extremal Transitions in }

{\large\bf Heterotic String Theory }

\vspace{1in}

Eric Sharpe \\
Physics Department \\
Princeton University \\
Princeton, NJ  08544 \\
{\tt ersharpe@puhep1.princeton.edu}

\vspace{0.75in}

\end{center}

In this paper we study extremal transitions between heterotic string
compactifications, i.e., transitions between pairs $(M,V)$ where $M$ is a
Calabi-Yau manifold and $V$ a gauge bundle.  Bundle transitions are
described using language recently espoused by Friedman, Morgan,
Witten.  In addition, partly as a check on our methods, we also study
how small instantons are described in the same language, and also 
describe the sheaves corresponding to small
instantons.

\begin{flushleft}
May 1997
\end{flushleft}

\newpage

\section{\label{intro} Introduction}

Historically one of the biggest challenges facing string theorists was
to understand the vacuum degeneracy problem, that is, why our universe
would be described by one particular string theory compactification,
out of the many possibilities.  It was known for some time that
Calabi-Yau's can be connected through a series of extremal transitions
\cite{reid,splittype,toricextreme},
but in the degeneration limits connecting
distinct Calabi-Yaus, the conformal field theory broke down, and so such
transitions were not believed to be realized physically.

This perspective was radically changed by the work of Strominger
\cite{strom} 
and Greene, Morrison, Strominger \cite{gms} who showed in detail
how nonperturbative effects would cure all such ills in type II
compactifications of string theory.

However, their work did not touch on the problem of relating distinct
heterotic compactifications.  To compactify heterotic string theory, one
must specify more than just a Calabi-Yau, one must also specify (at least)
one vector bundle (or, more generally, a sheaf) which breaks $E_{8}
\times E_{8}$ or $Spin(32)/{\bf Z}_{2}$ to a subgroup.  Although the space of
Calabi-Yaus may be connected, one must also understand how the vector
bundle changes during the transition.

Some amount of light was shed on this question by F-theory
compactifications \cite{fthy,btate,paulgross}.  
Compactifications of F-theory on an
elliptic 
Calabi-Yau $n$-fold have a weak coupling limit in which they can be
described as compactifications of heterotic string theory on a
Calabi-Yau $(n-1)$-fold, with a gauge bundle implicitly specified 
in the form of the elliptic fibration of the F-theory compactification.
Extremal transitions between F-theory compactifications have been
discussed recently in \cite{modsplit,modblow}.

Unfortunately F-theory compactifications are only dual to heterotic
compactifications on elliptic fibrations.  Heterotic string theory can,
of course, be compactified on much more general Calabi-Yaus, so not all
heterotic extremal transitions can be understood within F-theory.  
In addition, F-theory compactifications often yield non-chiral heterotic
duals (in the sense that the Dirac index of the vector bundle vanishes),
unless one turns on background fields (for a few comments on this issue
see section 7.1 of \cite{wmf}).  As is well-known, the Dirac index is
invariant under smooth deformations of the theory (modulo potentially
exciting IR dynamics, as in \cite{sechir}), so in particular it
should remain invariant through an extremal transition.  One would like
to understand extremal transitions for arbitrary heterotic
compactifications, in which the Dirac index does not necessarily vanish.
Thus, in order to understand extremal transitions between non-elliptic
heterotic compactifications, and to have a less cumbersome method of
understanding chiral heterotic compactifications, it is not sufficient
to work within F-theory.

In this paper, we develop methods to describe extremal transitions 
directly in heterotic string
theories, using technology
advanced very recently by Friedman, Morgan, Witten \cite{wmf}, and in
related work by \cite{bbundle,donagi}, and also \cite{bs,sechir,acl,pauldave}.  
As a result, we only consider heterotic
compactifications on elliptic varieties -- in a companion paper
\cite{me} we will study extremal transitions using the (0,2) models of
Distler, Kachru \cite{(02)}, which are not constrained to elliptic
varieties.  In addition, to better understand the
relation between bundle degenerations and nonperturbative physics, we also 
study the sheaves
associated with small instantons on $K3$, and the corresponding spectral cover
degenerations.  

A weakness of the present work is that we have relatively little
to say about the precise nonperturbative physics occurring in the
degeneration limits.  For example, any potential extremal transition
described in this paper, which has constant Dirac index through the
transition, could be obstructed by a
spacetime superpotential.  In transitions which are not obstructed, we expect that
in many cases asymptotically-free N=1 supersymmetric gauge dynamics will 
account for singular behavior in the degeneration limit; for an
example in which such behavior is studied, see \cite{ses}.

Another weakness is that we will be ignoring potential worldsheet
instanton effects.  Not only are we working classically in string loops,
but also classically on the worldsheet. 

This paper divides naturally into two parts.  After a review of
constraints on heterotic compactifications in section two, in section
three we make a detailed study of how a gauge bundle on an elliptic
three-fold with base ${\bf F}_{1}$ transforms under an extremal
transition to a three-fold with base ${\bf P}^{2}$.  After reviewing the
geometry of this extremal transition, we study how the
bundle transforms, then check our results by studying the transformation
of the spectral
cover defining the bundle.  The second part of the paper is in section
four, in which we study vector bundle degenerations over $K3$.  We work 
out details of the sheaf corresponding to small
instantons on $K3$, and we study the corresponding spectral curve
degeneration.  We conclude in section five, and have also
included
a pair of appendices, on the basics of ideal sheaves and homological
algebra.

\section{A Rapid Review of Heterotic Compactifications}

Before studying extremal transitions between distinct heterotic
compactifications, we will first review some basics of such compactifications.

For a consistent perturbative compactification of either the
$E_{8}\times E_{8}$ or $Spin(32)/{\bf Z}_{2}$ heterotic string, in addition to
specifying a Calabi-Yau $Z$ one must also specify 
a set of stable \cite{okonek},
holomorphic vector bundles (or, more generally,
sheaves)
$V_{i}$.
These vector bundles must obey certain constraints.  For $U(n)$
bundles\footnote{Strictly speaking, in this paper we will consider
bundles whose structure group is the
complexification of $U(n)$, but this should cause minimal confusion.}
one constraint can be written as
\begin{displaymath}
\omega^{n-1} \cup c_{1}( V_{i} ) \: = \: 0
\end{displaymath}
where $n$ is the complex dimension of the Calabi-Yau, and $\omega$ is
the Kahler form, and the other is an anomaly-cancellation condition which, 
if a single $V_{i}$ is embedded in each $E_{8}$, can be written as
\begin{displaymath}
\sum_{i} \, \left( c_{2}(V_{i}) \, - \, \frac{1}{2} c_{1}(V_{i})^{2} \,
\right) \: = \: c_{2}(TZ)
\end{displaymath} 

It was noted \cite{dmw} 
that the anomaly-cancellation conditions can be
modified slightly by the presence of five-branes in the heterotic
compactification.  Let $[W]$ denote the cohomology class of the
five-branes, then the second constraint above is modified to
\begin{displaymath}
\sum_{i} \, \left( c_{2}(V_{i}) \, - \, \frac{1}{2} c_{1}(V_{i})^{2} \,
\right) \: + \: [W] \: = \: c_{2}(TZ)
\end{displaymath}

Historically, for a long time the only perturbative heterotic
compactifications studied were those in which one took $V \, = \, TZ$,
the ``standard embedding."
This was done partly because more general compactifications are 
more difficult to work with, and partly because it was believed more
general compactifications were destabilized by worldsheet instantons
\cite{xen}.  For perturbative compactifications described by
gauged linear sigma models, both difficulties have been overcome
\cite{(02),eva(02)}. 

When does one expect nonperturbative effects in heterotic string theory?
For perturbative compactifications, one will get nonperturbative effects
when the CFT breaks down. 
At least for CFTs which are
low-energy limits of gauged linear sigma models \cite{phases,(02)},
such degenerations are controlled entirely by the vector
bundle, and not at all by the Calabi-Yau base space, surprisingly
enough.  This would appear to contradict a naive argument from M theory:
Consider the heterotic string in a strong coupling limit in which it is
described as M theory compactified on $(S^{1}/{\bf Z}_{2}) \times X$, where $X$
is some Calabi-Yau space.  Now, in M theory compactifications on
$X$, 
a degeneration
of $X$ signals the occurrence of nonperturbative effects.  Naively
one would expect the same to be true for M theory on $(S^{1}/{\bf Z}_{2})
\times X$.
However, the massless particles in bulk are lifted by boundary
effects, which is possible because the boundary theory has no BPS
states\footnote{For 
$X$ a Calabi-Yau three-fold, the boundary theory has N=1 supersymmetry
in four dimensions, which has no BPS states.  For $X$ a $K3$, the
boundary theory has chiral (1,0) supersymmetry in six dimensions, and
chiral superalgebras do not admit central charges.}.

\section{Extremal Transitions and Spectral Covers}

In this section we will consider a prototypical example of an extremal
transition which preserves an elliptic fibration structure -- an extremal
transition between an elliptic three-fold fibered over ${\bf F}_{1}$,
and an elliptic three-fold fibered over ${\bf P}^{2}$.
Given an $SU(n)$ vector bundle $V$ over an elliptic three-fold with base ${\bf
F}_{1}$, constructed following \cite{wmf}, we hypothesize that after the
three-fold is transformed into an elliptic fibration over ${\bf P}^{2}$,
we will get a sheaf which can be deformed into another $SU(n)$ vector bundle.
We argue that the sheaf appearing in the degenerate limit is locally
free\footnote{For readers less well acquainted with sheaf theory, a few
definitions are in order.  In this paper, by ``locally free" we mean a
sheaf that is associated to a well-defined vector bundle.  If a sheaf is
locally free everywhere except along some codimension two subvariety,
then we refer to it as ``torsion-free" -- i.e., a torsion-free sheaf
looks like a vector bundle except on some codimension two subvariety
where the bundle degenerates.  For more information see for example
\cite{okonek}.} away from a codimension two locus, which is naturally identified
with small instantons / five-branes that necessarily appear in order to
satisfy anomaly cancellation.  We check that this is consistent with the
description of the bundle in terms of spectral covers, by observing that
the spectral cover of $V$ transforms naturally under the blowdown into
the spectral cover of another $SU(n)$ bundle, modulo singular behavior
at the codimension two locus just mentioned.  Bundles over related spaces have also been considered in \cite{acl}, though with a different perspective. 

\subsection{Review of Three-Fold Geometry}

We will begin by briefly reviewing the geometry of the extremal 
transition between elliptic fibrations over ${\bf F}_{1}$ and ${\bf
P}^{2}$, as has been previously discussed in \cite{fthy,kmv}.

As is well-known, ${\bf F}_{1}$ can be viewed as a blowup of ${\bf
P}^{2}$, so in order to send ${\bf F}_{1} \rightarrow {\bf P}^{2}$, one
need merely blow down the exceptional curve of self-intersection $-1$. 

Naively one might think that to transform a three-fold elliptically
fibered over ${\bf F}_{1}$ into one elliptically fibered over ${\bf
P}^{2}$, one need merely blow down a divisor containing the exceptional
curve in ${\bf F}_{1}$.  In the limit that the radius of the elliptic
fiber is zero (the limit relevant for F theory compactifications on this
three-fold), this is indeed the case.  But unfortunately for nonzero
Kahler modulus, the full story is somewhat more complicated.

At nonzero Kahler modulus of the fiber, the transition between
three-folds is a two-step process.  First, one must perform a flop in
the three-fold.  Before the flop, the elliptic fibration over ${\bf
F}_{1}$ is a $K3$-fibration, and the divisor containing the exceptional
curve of ${\bf F}_{1}$ is a rational elliptic surface (${\bf P}^{2}$ with nine blowups).
The flop acts by shrinking the section of the rational elliptic surface
divisor to a point, then inserting another ${\bf P}^{1}$ at this point
and orthogonal to the divisor, as sketched in Figure \ref{flop}.
After the flop, the rational elliptic surface becomes 
${\bf P}^{2}$ with eight blowups. 
The ``intermediate"
three-fold looks mostly like an elliptic fibration over ${\bf P}^{2}$,
with the exception of a four-cycle (the ${\bf P}^{2}$ with eight blowups
just mentioned).  Finally, one blows down this four-cycle (the
${\bf P}^{2}$ with eight blowups) to recover a three-fold that is
globally an elliptic fibration over ${\bf P}^{2}$.  The ${\bf P}^{1}$
created during the flop becomes a (singular) elliptic fiber after
blowing down the four-cycle.

The elliptic fibration over ${\bf P}^{2}$ that one obtains via the
birational transformation above is singular, and the
singularity must be resolved by deformation of complex structure
(as opposed to a blowup of the elliptic fiber).

\subsection{General Remarks on Transforming the Bundle}

To make this exercise as straightforward as possible\footnote{And
for other reasons.  According to \cite{friedprivate,bbundle}, the calculation of
$c_{3}$ in the language of spectral covers is somewhat subtle,
essentially because for a three-fold $X$, the fibered product $X
\times_{B} \Sigma$, $\Sigma$ the spectral cover, is necessarily
singular.}, we will
restrict to $SU(n)$ bundles of $c_{3} \, = \, 0$.  Therefore the Dirac
index of the bundle is automatically constrained to be constant through the
transition.

We will push the locally free sheaf associated to the bundle on the
three-fold over ${\bf F}_{1}$ through the flop and the divisor blowdown,
using pushforwards and pullbacks.

\begin{figure}
\centerline{\psfig{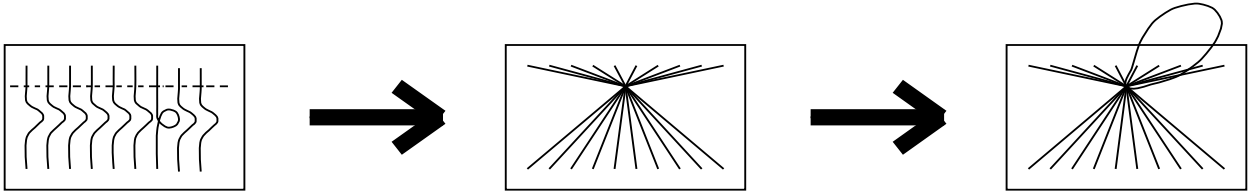}}
\caption{\label{flop} Transformation of the rational elliptic surface
divisor under the flop.  }
\end{figure}

First, consider the first half of the flop, in which a ${\bf P}^{1}$ is
shrunk to a point.  For definiteness, let $E$ denote the exceptional
curve in ${\bf F}_{1}$, and let $p \in {\bf P}^{2}$ denote the point to which the
exceptional curve $E$ shrinks in the blowdown ${\bf F}_{1} \rightarrow
{\bf P}^{2}$.  In the first half of the flop, the ${\bf P}^{1}$ that
shrinks to a point is the image of $E$ in the section of the rational
elliptic surface, 
as shown in Figure \ref{flop}.  Let $\pi_{1}$ denote this morphism.
If $V$ is the locally free sheaf associated to a vector bundle over
the elliptic three-fold with base ${\bf F}_{1}$, then after the first
half of the flop, $V \rightarrow \pi_{1 *} V$.

How can we compute $\pi_{1 *} V$ ?  Clearly it is sufficient to restrict
to the shrinking ${\bf P}^{1}$.  As was shown by Grothendieck, all
holomorphic vector bundles over ${\bf P}^{1}$ split, so when restricted
to ${\bf P}^{1}$, $V$ will have the form
\begin{displaymath}
{\cal O}(n_{1}) \oplus {\cal O}(n_{2}) \oplus {\cal O}(n_{3}) \oplus
\cdots
\end{displaymath}
and since $c_{1}(V)\, = \, 0$, $\sum n_{i} \, = \, 0$.  If $n \geq 0$,
then $\pi_{1 *} {\cal O}(n)$ is a skyscraper sheaf of rank $n+1 \, = \,
h^{0}({\bf P}^{1},{\cal O}(n))$.  If $n < 0$, then $\pi_{1 *} {\cal O}(n)$
is an ideal sheaf, vanishing to order $-n$.

Now consider the second half of the flop.  Let $\pi_{2}$ denote the
morphism projecting the new ${\bf P}^{1}$ back down to the (singular)
point we reached at the end of the first half of the flop.  To pull the
sheaf through, compute $\pi_{2}^{*} \pi_{1 *} V$.\footnote{Given a
morphism of varieties $f: Y \rightarrow X$ and a sheaf $V$ over $X$, there are other
ways to pull the sheaf $V$ back to a sheaf over $Y$ than $f^{*} V$ --
for example, one could compute $f^{-1} V$.  However, $f^{-1}$ of a
locally free sheaf is not necessarily locally free, so we will only
consider $f^{*} V$.}  The pullback of a rank $n$ skyscraper sheaf with
support over the insertion point will be
a skyscraper sheaf with support on the ${\bf P}^{1}$, and on its support
will have the form ${\cal O}_{{\bf P}^{1}}^{\oplus n}$.  The pullback
of an ideal sheaf which vanishes at the insertion point is a
sheaf which vanishes to the same order as at the insertion point, along
the entire exceptional divisor.

So far we have formally pushed $V$ through the flop using pushforwards and
pullbacks, and we have outlined how to see that the sheaf fails to be
locally free along the ${\bf P}^{1}$ created during the flop.  (Note,
incidentally, that since the sheaf fails to be locally free at
codimension two, it is always torsion-free.)  In passing we should note
that this procedure does not uniquely identify the sheaf that may appear
after the flop -- for a recent discussion of this issue, see for
example \cite{brussee}.

Now, all that remains is to blowdown a divisor (${\bf P}^{2}$ with eight
blowups) in order to recover an elliptic fibration over ${\bf P}^{2}$.
Here again, if $\pi$ denotes the blowdown morphism, then we need merely
compute $\pi_{*} \pi_{2}^{*} \pi_{1 *} V$.  

The image of $V$ after blowing down the divisor will be locally free
everywhere except along the elliptic fiber over $p \in {\bf P}^{2}$.
This singular elliptic fiber is precisely the image of the ${\bf P}^{1}$
created during the flop.  We will show later by a simple counting
argument that in this transition, some five-branes appear, wrapped on
precisely this elliptic fiber -- so the failure of local freedom of the
sheaf
coincides with the location of the five-branes. 

\subsection{Analysis of the Spectral Cover}

Having made these general observations, we now turn to a detailed
analysis of the spectral cover.  As the spectral cover is defined over
precisely the base of the elliptic fibration, we will be able to largely
ignore the complexities of the three-fold geometry of the transition.
We will discover that the spectral cover degenerates at $p \in {\bf
P}^{2}$, in precise agreement with our general remarks above on failure
of local freedom of the sheaf along the elliptic fiber over this point.
We will also discover that the spectral cover of an $SU(n)$ bundle
transforms into a (singular) spectral cover of another $SU(n)$ bundle,
which can be deformed to describe a rank $n$ vector bundle, i.e., the
rank is unchanged by the extremal transition.

We begin by describing bundles on three-folds with base ${\bf
F}_{1}$.  (Then, we will blowdown ${\bf F}_{1} \rightarrow {\bf P}^{2}$
and describe what happens to the bundle.)
The Hirzebruch surface can be described by 
a GIT\footnote{Geometric Invariant Theory.  For more information,
see \cite{mumford}.  In this paper we will be sloppy and usually ignore
the fact that, in addition to specifying ${\bf C}^{\times}$ actions on a
set of homogeneous coordinates, one must also specify a set of points to
be omitted before quotienting.} quotient, i.e., homogeneous coordinates and ${\bf
C}^{\times}$ actions, as
\begin{center}
\begin{tabular}{c|cccc}
 \, & $u$ & $v$ & $w$ & $s$ \\ \hline
$\lambda$ & 1 & 1 & 1 & 0 \\
$\mu$ & 0 & 0 & 1 & 1 \\
\end{tabular}
\end{center}
In other words, just as a projective space can be described in terms of
homogeneous coordinates identified under single ${\bf C}^{\times}$
action, here we are describing ${\bf F}_{1}$ in terms of homogeneous
coordinates $u$, $v$, $w$, and $s$, identified under two ${\bf
C}^{\times}$ actions $\lambda$, $\mu$ with weights as shown above.

In this description we can recover ${\bf P}^{2}$ (the blowdown limit) by
using the ${\bf C}^{\times}$ action $\mu$ to fix
$s$ to a nonzero value, then $u$, $v$, $w$
are homogeneous coordinates on ${\bf P}^{2}$, and the blowup is located
at $u \, = \, 0, \: v \, = \, 0$ (the point $p \in {\bf P}^{2}$).

We will assume there are two unitary gauge bundles, $V_{1}$, $V_{2}$,
each of $c_{1} \, = \, 0$ and of ranks $n_{1}$, $n_{2}$, respectively.
Let $Z$ denote an elliptic three-fold fibered over ${\bf F}_{1}$, i.e., $\pi: \, Z \,
\rightarrow \, {\bf F}_{1}$.  Following closely the notations and conventions of
\cite{wmf}, we will only demand $\pi_{*} c_{2}(V_{1}) \: + \: \pi_{*}
c_{2}(V_{2}) \: = \: \pi_{*}
c_{2}(TZ)$, both before and after the transition ${\bf F}_{1}
\rightarrow {\bf P}^{2}$, rather than the full condition for perturbative
compactifications $c_{2}(V_{1}) \: + \: c_{2}(V_{2}) \: = \: c_{2}(TZ)$, so the background 
will contain five-branes, wrapped on the elliptic fiber.  

The spectral cover of an $SU(n)$ bundle is specified by functions
$a_{0}$, $a_{2}$, $a_{3}$, \ldots, which are sections of the bundles
${\cal N}$, ${\cal N} \otimes {\cal L}^{-2}$, ${\cal N} \otimes {\cal
L}^{-3}$, \ldots, where ${\cal L} \, = \, K_{{\bf F}_{1}}^{-1}$, 
and ${\cal N}
\rightarrow {\bf F}_{1}$ is a line bundle.  
Also, recall from \cite{wmf} that $\pi_{*}
c_{2}(TZ) \: = \: 12 \, c_{1}({\bf F}_{1})$.  Finally, letting ${\cal
N}_{i}$ denote the bundle associated with the spectral data of $V_{i}$,
for reasons of technical convenience we will impose a pair 
of additional conditions\footnote{These conditions impose a particular
${\bf Z}_{2}$ symmetry on the bundle.  See equation (7.49) of
\cite{wmf}.} on the bundle: 
\begin{eqnarray*}
n_{i} & \equiv & 0 \mbox{ mod 2, for all $i$} \\
c_{1}({\cal N}_{i}) & \equiv & c_{1}({\cal L}) \mbox{ mod 2, for all
$i$}
\end{eqnarray*}
and for illustration we will demand $c_{1}({\cal N}_{1}) \, = \, (6+t) c_{1}({\cal L})$,
$c_{1}({\cal N}_{2}) \, = \, (6-t) c_{1}({\cal L})$, for some odd
integer $t$.

Before we go on, we will introduce a little notation.  Let $D_{u}$
denote both the divisor $\{ u \, = \, 0 \}$ and its Poincare-dual
element of $H^{2}({\bf F}_{1},{\bf Z})$, in somewhat sloppy notation.  (So
in particular if we view ${\bf F}_{1}$ as a ${\bf P}^{1}$ fibration,
$D_{u}$ is the fiber and $D_{s}$ is the isolated section.)  Let ${\cal O}(m,n)$ denote the
line bundle over ${\bf F}_{1}$ such that $c_{1} \, = \, m D_{u} \, + \,
n D_{s}$, so in particular $c_{1}({\bf F}_{1}) \, = \, 3 D_{u} \, + \, 2
D_{s}$.  

Putting this all together, we find
\begin{eqnarray*}
{\cal L} & = & {\cal O}(3,2) \\
{\cal N}_{1} & = & {\cal L}^{6+t} \\
{\cal N}_{2} & = & {\cal L}^{6-t} \\
a_{0}(1) & \in & \Gamma({\cal N}_{1}) \: = \: \Gamma({\cal O}(18+3t,12+2t)) \\
a_{2}(1) & \in & \Gamma({\cal N}_{1}\otimes{\cal L}^{-2}) \: = \: \Gamma({\cal O}(12+3t,8+2t)) \\
a_{3}(1) & \in & \Gamma({\cal N}_{1}\otimes{\cal L}^{-3}) \: = \: \Gamma({\cal O}(9+3t,6+2t)) \\
a_{4}(1) & \in & \Gamma({\cal N}_{1}\otimes{\cal L}^{-4}) \: = \: \Gamma({\cal O}(6+3t,4+2t)) 
\end{eqnarray*}
and so forth.

We can expand out the $a_{k}$ in terms of the homogeneous coordinates on
${\bf F}_{1}$ as follows:
\begin{eqnarray*}
a_{0}(1) & = & \sum_{i,j} \, a_{0,i,j} \, u^{i} \, v^{j} \, w^{18+3t-i-j} \,
s^{i+j-6-t} \\
a_{2}(1) & = & \sum_{i,j} \, a_{2,i,j} \, u^{i} \, v^{j} \, w^{12+3t-i-j} \,
s^{i+j-4-t} \\
a_{3}(1) & = & \sum_{i,j} \, a_{3,i,j} \, u^{i} \, v^{j} \, w^{9+3t-i-j} \,
s^{i+j-3-t} \\
a_{4}(1) & = & \sum_{i,j} \, a_{4,i,j} \, u^{i} \, v^{j} \, w^{6+3t-i-j} \,
s^{i+j-2-t} 
\end{eqnarray*}
and so forth.  

Were we on ${\bf P}^{2}$ rather than ${\bf F}_{1}$, the expansions of
the $a_{k}$ in terms of homogeneous coordinates would be identical,
except for the omission of the $s$ factor. 

Note that the $s$ factor constrains the $a_{k}$ on ${\bf F}_{1}$ more than they would be
on ${\bf P}^{2}$.  To be specific, consider $a_{0}$(1).  On ${\bf P}^{2}$, the sum over $i,j$
would run over $0 \, \leq \, i+j \, \leq 18+3t$, whereas on ${\bf F}_{1}$ because of the $s$
factor the sum is restricted to $6+t \, \leq \, i+j \,
\leq \, 18+3t$.  In particular this means that each of the $a_{k}(1)$ (for $k
\leq 5+t$) vanishes over the point $p \in {\bf P}^{2}$ ($u \, = \, v \,
= \, 0$),
so in the blowdown limit of ${\bf F}_{1}$ the vector bundle 
becomes some torsion-free sheaf\footnote{Judging from the codimension of
the singularities in the bundle.  In general, the fact that
the $a_{k}$ all vanish at a point does not necessarily imply the
bundle degenerates over that point -- this will be discussed in
section \ref{one5brane}.}.  Note that this is consistent with
the remarks made earlier, that after pulling $V$ through the birational
transformation it fails to be locally free on the (singular) elliptic
fiber over $p \in {\bf P}^{2}$.

The idea in the last paragraph will appear many more times in this paper
and so is worth repeating.  Given a section of some line bundle over
$\tilde{X}$, a blowup of $X$, in the blowdown limit the section will
often have zeroes at the location of the blowup on $X$.  By
expanding out sections of line bundles explicitly in terms of
homogeneous coordinates, we are able to pick off this behavior
directly.

There is a more invariant method to
describe this result.  In general, suppose $\pi: \tilde{X} \rightarrow
X$ is a blowup of $X$, and $\pi^{*}(L) \otimes {\cal O}(-n)$ some line bundle
over $\tilde{X}$, where 
$c_{1}({\cal O}(-n))$ is $-n$ times the dual to the
exceptional divisor.  Then the direct image sheaf $\pi_{*}(\pi^{*}(L) \otimes
{\cal O}(-n)) \: = \: L \otimes {\cal I}_{n}$, where ${\cal I}_{n}$ is an ideal sheaf
on $X$
which vanishes to order $n$ at the location of the blowup.  In the
present case, consider for example $a_{0}$(1).  Since $a_{0}(1) \, \in \,
\Gamma({\cal N}_{1})$ and $c_{1}({\cal N}_{1}) \, = \, 3(6+t) u_{0} \, - \,
(6+t) 
u_{1}$ (in conventions where $K_{{\bf F}_{1}} \, = \, -3 u_{0} + u_{1}$,
so $u_{1}$ is dual to the exceptional divisor of the blowup of ${\bf
P}^{2}$), we have $\pi_{*}({\cal N}_{1}) \, = \, K_{{\bf P}^{2}}^{-(6+t)} \otimes {\cal I}_{6+t}$,
and indeed we showed in the last paragraph that in the blowdown limit,
$a_{0}(1)$ vanishes to order $6+t$ at $p \in {\bf P}^{2}$.

At this point we will take a moment to describe how the complex
structure of the elliptic
fibration $Z \rightarrow {\bf F}_{1}$ degenerates, as another example of
the concept above.
$Z$ is given as a
hypersurface in the total space of ${\bf P}({\cal O} \oplus {\cal L}^2
\oplus {\cal L}^{3}) \, \rightarrow \, {\bf F}_{1}$, with homogeneous
coordinates $(z,x,y)$ (respectively) on the ${\bf P}^{2}$ fiber.
The hypersurface is of the form $y^{2} z \: = \: x^{3} \: + \: f x z^{2} \:
+ \: g z^{3}$, where $f \, \in \, \Gamma({\cal L}^{4})$, $g \, \in \,
\Gamma({\cal L}^{6})$.  Since $c_{1}({\cal L}) \, = \, 3 u_{0} \, - \,
u_{1}$, we can read off that in the blowdown limit, $f$ will have an order
$4$ zero, $g$ an order $6$ zero, at $p \in {\bf P}^{2}$.  Clearly we
can resolve this singularity in $Z \rightarrow {\bf P}^{2}$ by deforming
the complex structure.  One can often also resolve singularities by
blowups of the elliptic fiber; however, in the present case, the singularity in $Z$ is too
severe to be blown up fiber-wise into another Calabi-Yau.  The bundle degeneration
can be resolved simply by deforming the $a_{k}$ to more generic
sections.

Conversely, consider starting with an $SU(n)$ bundle on $Z \rightarrow {\bf P}^{2}$.
In order to blow up the point $p \in {\bf P}^{2}$, we must first adjust
both the sections defining the Weierstrass fibration as well as the
$a_{k}$ defining the bundle.  In particular, it is not sufficient to
arrange for only the base to be singular -- in order to be able to blow
up the base with bundle consistently, one must also adjust the $a_{k}$
to be singular.  (This is in accordance with the observation in
the introduction that, for perturbative heterotic
compactifications, the conformal field theory degeneration is controlled
by the vector bundle -- so deforming only the base to the singular locus is
insufficient for the conformal field theory to break down, and make an
extremal transition possible.) 

So far we have described how in a transition between elliptic
three-folds over ${\bf F}_{1}$ and ${\bf P}^{2}$, the $a_{k}$ defining
the spectral cover for a pair of $SU(n)$ bundles changes.  To completely
specify the bundles, one must in addition specify a line bundle on each
spectral cover.  We assumed at the beginning that each bundle was
invariant under a ${\bf Z}_{2}$ symmetry (called $\tau$ in \cite{wmf}),
in which case the line bundle on the spectral cover is trivial.  In
particular, since the bundles on either side of the transition are taken
to be $\tau$ invariant, the line bundles on the spectral covers are
trivial throughout the transition.  

How many five-branes appear in the transition?
Recall that if $[W]$ denotes the cohomology class of the five-branes,
then
\begin{eqnarray*}
[W] & = & c_{2}(TZ) \: - \: \sum_{i} \, c_{2}(V_{i}) \\
 \, & = & 11 \, c_{1}({\cal L})^{2} \: + \: c_{2}(B) \: - \:
\frac{1}{24} \, c_{1}({\cal L})^{2} \, \left[ \, n_{1}^{3} \, - \, n_{1} \,
+ \, n_{2}^{3} \, - \, n_{2} \, \right] \\
\, & \, & - \: \frac{1}{8} \, c_{1}({\cal L})^{2} \, \left[ \, n_{1} \,
(6+t) \, (6+t-n_{1}) \, + \, n_{2} \, (6-t) \, (6-t-n_{2}) \, \right]
\end{eqnarray*}
where $B$ is the base of the elliptic three-fold (either ${\bf F}_{1}$
or ${\bf P}^{2}$).
Despite appearances, it can be shown using the mod 2 conditions
mentioned earlier that $[W]$ is an element of integral cohomology, that
is, that the number of five-branes is an integer.
To insure supersymmetry is unbroken one must check in general that the
number of five-branes is nonnegative.  In any event, it is clear that in
general the number of five-branes present (wrapped on the elliptic
fiber) is different on either side of the transition.  Recall from our
discussion above that in the blowdown ${\bf F}_{1} \rightarrow {\bf
P}^{2}$, the bundle becomes singular along the singular elliptic fiber
over $p \in {\bf P}^{2}$;
clearly this codimension two locus
should be interpreted as the locations of five-branes.

To review, given a pair of $\tau$-invariant bundles on an elliptic
three-fold fibered over ${\bf F}_{1}$, we have explicitly worked out how
the bundles transform under the blowdown ${\bf F}_{1} \rightarrow {\bf
P}^{2}$, in terms of the spectral data defining them.  In the degenerate
limit the bundle has a singularity over a codimension two locus, which
is interpreted as due to the presence of five-branes.
In principle the same
idea should apply much more generally.  (Although not all rational
surfaces are toric, we have outlined how to attack other cases, in terms
of direct image sheaves and ideal sheaves).

We have not attempted to determine whether the spacetime superpotential
obstructs this transition, though in principle it might be possible to
work this out following \cite{witsup,dgw}.

\subsection{Splitting-Type Extremal Transitions}

Lest the reader get the impression that all bundle degenerations are
reflected by a spectral cover degeneration, in this section we will give
a counterexample.  We will consider a splitting-type transition \cite{splittype,modsplit},
between an elliptic three-fold with base ${\bf P}^{2}$ and another
elliptic three-fold with the same base.  It will turn out that at the
transition point, singularities in the elliptic three-fold will lie
along the section of the elliptic fibration.  We will not have anything
to say about how bundles behave through such a transition, 
nevertheless we felt it appropriate to include this
discussion.

First we will describe an elliptic three-fold with base ${\bf P}^{2}$.
The Calabi-Yau is described as a hypersurface in an ambient space
which is obtained by fibering ${\bf P}^{2}$ over the base ${\bf P}^{2}$.
Let $u$, $v$, $w$ be homogeneous coordinates on the base and $x$, $y$,
$z$ homogeneous coordinates on the fiber, then we have ${\bf
C}^{\times}$ actions as
\begin{center}
\begin{tabular}{c|ccc|ccc} 
\, & $u$ & $v$ & $w$ & $x$ & $y$ & $z$ \\ \hline
$\lambda$ & 1 & 1 & 1 & $2 \alpha$ & $3 \alpha$ & 0 \\
$\nu$ & 0 & 0 & 0 & 1 & 1 & 1 \\
\end{tabular}
\end{center}
with hypersurface defined by
\begin{displaymath}
y^{2} z \: = \: x^{3} \: + \: f(u,v,w) x z^{2} \: + \: g(u,v,w) z^{3} 
\end{displaymath}
For this hypersurface to be Calabi-Yau, we demand $\alpha \, = \, 3$.
Also, $f$ has degree $4 \alpha \, = \, 12$ under $\lambda$,
and $g$ has degree $6 \alpha \, = \, 18$ under $\lambda$.

For convenience, let $M$ denote the ambient space described above.
Consider a splitting-type transition in which the hypersurface $W \, = \, 0$ in
$M$ transforms into a complete intersection $W_{1} \, = \, W_{2} \, = \,
0$ in ${\bf P}^{1} \times M$, and the degrees of the
hypersurfaces are
\begin{displaymath}
\begin{array}{c|cc|}
{\bf P}^{1} & 1 & 1 \\
\lambda & 6 & 12 \\
\nu & 1 & 2
\end{array}
\end{displaymath}

Note that the complete intersection above has base ${\bf P}^{2}$, the
same as previously, and fiber
\begin{displaymath}
\begin{array}{c|cc|}
{\bf P}^{1} & 1 & 1 \\
\nu & 1 & 2 
\end{array}
\end{displaymath}
which is manifestly an elliptic curve.  So, in other words, this
particular splitting-type transition takes place entirely within the
elliptic fiber.  In addition, on both sides of the transition the
elliptic fibration has a section:  the threefold in M has elliptic
section $\{ \, x \, = \, z \, = \, 0, \: y \, = \, 1 \, \}$, and the
threefold in $\mbox{M} \times {\bf P}^{1}$ has section $\{ \, x \, = \,
z \, = \, 0, \: y \, = \, 1, \: t_{0} \, = \, 0, \: t_{1} \, = \, 1 \,
\}$ where $t_{0}, \, t_{1}$ are homogeneous coordinates on the ${\bf
P}^{1}$. 

The fact that the base is invariant under this transformation somewhat
simplifies the analysis of vector bundles over the three-fold in the
language of \cite{wmf}.  Suppose, for definiteness, we have an $SU(n)$
bundle, whose spectral cover is specified by $a_{k}$ (sections of
line bundles over the base).  Since the base of the elliptic fibration
is unchanged by the transition, and the $a_{k}$ are sections of bundles
over the base, the $a_{k}$ are invariant through the transition.

Although the $a_{k}$ are invariant through the transition, the
description of the bundle does break down at the transition point,
because all of the conifold singularities are located along the section of the
elliptic fibration.  This is relatively straightforward to see.  Let
$W_{1}$, $W_{2}$ be the hypersurfaces in $\mbox{M} \times {\bf P}^{1}$
whose complete intersection is the Calabi-Yau.  Write
\begin{eqnarray*}
W_{1} & = & t_{0} \, P \: + \: t_{1} \, Q \\
W_{2} & = & t_{0} \, R \: + \: t_{1} \, S
\end{eqnarray*}
then in the blowdown limit, this complete intersection becomes the
hypersurface $P S \, - \, Q R \, = \, 0$, with (conifold) singularities
at $P \, = \, Q \, = \, R \, = \, S \, = \, 0$.  By expanding out $P$,
$Q$, $R$, and $S$, it is easy to see they all vanish at $x \, = \, z \,
= \, 0$, $y \, = \, 1$, which is precisely the section of the elliptic
fibration.

Clearly the bundle degeneration that occurs in this transition is quite
different from that in the previous section.  There, the bundle
singularity was signaled by the fact that the $a_{k}$ all vanished over
some point on the base, at the transition point.  Here, by contrast,
the $a_{k}$ are unaffected by the transition!

\section{\label{nonpertdict}  Vector Bundle Degenerations on Surfaces}

In this section we examine vector bundle degenerations over $K3$.  We
begin by analyzing the case of a single small instanton, by conjecturing
the form of the sheaf describing such a degeneration and then by using
F theory to understand the precise spectral cover behavior.  
We also examine
spectral cover degenerations corresponding to
multiple small instantons.

\subsection{Small Instantons and Sheaves on $K3$}

In this subsection we conjecture that the precise
sheaf corresponding to a small instanton of $SU(2)$ is of
the form ${\cal O} \oplus {\cal J}$ in a neighborhood of the small
instanton, where ${\cal J}$ is an appropriate ideal sheaf.  
We motivate this conjecture by closely examining a
well-known \cite{sduality,vafainst} construction of $SU(2)$ bundles 
on $K3$ from an unordered set of points
on $K3$.  The arguments are necessarily rather technical in nature;
readers not familiar with
sheaf-theoretic homological algebra are encouraged to skip to the next
subsection.

First we shall review how to associate some number of unordered points
with an $SU(2)$ bundle $E$ on $K3$ of $c_{2}(E) \, = \, k$, closely following
\cite{sduality}.  First, find a line bundle $L$ on $K3$ such that
$\chi(E \otimes L^{-1}) = 1$, then generically $E \otimes L^{-1}$ will
have a unique (up to scalar multiple) section $s$.  
The section $s$ will have $c_{2}(E \otimes
L^{-1}) = 2k-3$ isolated zeroes.  The zeroes of this section are
precisely the unordered points that we associate with the bundle $V$.

Now, in order to make our conjecture regarding small instanton sheaves,
we shall closely examine the converse:  given a set of $2k-3$ points on
$K3$, we will construct a bundle $E$.   We construct $E$ as a nontrivial
extension \cite{gh,donkron}
\begin{displaymath}
0 \: \rightarrow \: L \: \rightarrow \: E \: \rightarrow \: L^{-1}
\otimes {\cal J} \:
\rightarrow \: 0
\end{displaymath}
where $L$ is an invertible sheaf and ${\cal J}$ is
an ideal sheaf, vanishing at the $2k-3$ points on $K3$.
These extensions are of course classified by
the group global $\mbox{Ext}^{1}(L^{-1} \otimes {\cal J}, L)$. 

The extensions $E$ are not necessarily locally free -- to recover a
bundle as an extension, one must impose additional constraints.  
Now, global $\mbox{Ext}$
is defined by a spectral sequence, and in particular
\begin{eqnarray*}
0 \: \rightarrow \: H^{0}(K3,{\it Ext}^{1}(L^{-1} \otimes {\cal J}, L)) \:
\rightarrow \: \mbox{Ext}^{1}(L^{-1} \otimes {\cal J}, L) & 
\rightarrow & H^{1}(K3, {\it Hom}(L^{-1} \otimes {\cal J}, L)) \\
\, & \, & \rightarrow \: H^{0}(K3, {\it Ext}^{2}(L^{-1}
\otimes {\cal J}, L))
\end{eqnarray*}
is an exact sequence.  (We are assuming the $K3$ has nonzero Picard
number, and that ${\cal J}$ vanishes at isolated points with
multiplicity one.)  Thus, elements of $\mbox{Ext}^{1}(L^{-1} \otimes {\cal
J}, L)$ are partly determined by elements of $H^{0}(K3, {\it
Ext}^{1}(L^{-1} \otimes {\cal J}, L)) \, = \, H^{0}(K3, {\cal O}/{\cal
J})$.  It can be shown \cite{gh} that the sheaf $E$ is locally free
precisely when each element of $H^{0}(K3, {\cal O}/{\cal J})$ is a unit. 

Thus, if we want to find non-locally-free sheaves that are in the same
S-equivalence classes as stable bundles on the moduli space, all we need
to do is pick extensions such that some element of $H^{0}(K3, {\cal
O}/{\cal J})$ is not a unit.  For example, if ${\cal O}/{\cal J}$ has
support over a point $x$ in $K3$, choose an extension corresponding to a
section with value $0$ over $x$.  Then, locally, the corresponding
extension will have the form ${\cal O} \oplus {\cal J}$ rather than
${\cal O} \oplus {\cal O}$.

This, then, is a possibility for the sheaf corresponding to small
instantons.  Simply, in a neighborhood in which the sheaf fails to be
locally free, it takes the form ${\cal O} \oplus {\cal J}$ (though is
not globally of this form) for some ideal sheaf ${\cal J}$.  Deforming the sheaf to be
locally free everywhere (by deforming the sections in a neighborhood of
$x$) would correspond to peeling a five-brane off the end-of-the-world,
in the language of M-theory compactified on $S^{1}/{\bf Z}_{2}$.  (As a
check, note that in such a deformation, $c_{2}$ drops by one.)

The reader may wonder why we did not mention another possibility:  that
when an instanton becomes small, the sheaf takes the form $V \otimes
{\cal I}$, where $V$ is locally free and ${\cal I}$ is an ideal sheaf
vanishing at the location of the small instanton.  A quick
Chern class computation will convince the reader that it is not possible
to describe only one small instanton\footnote{Though in principle such a
sheaf might describe multiple small instantons.}.

\subsection{\label{one5brane} Spectral Cover Degenerations on $K3$}

In this subsection we will work out the precise spectral cover
degeneration corresponding to a single small instanton on $K3$,
more precisely, a small $E_{8}$ instanton.  We will find that the
spectral cover becomes reducible, with one component being precisely
the spectral cover of a bundle with one less instanton.

We will use the duality between
heterotic strings on $K3$ and F theory on elliptic three-folds.
F theory on an elliptic three-fold (with section) whose
base is ${\bf F}_{n}$ is dual to a
heterotic compactification on $K3$ with instanton numbers $(12+n,12-n)$
embedded in either $E_{8}$.  Blowups of the base of the three-fold
correspond to instantons degenerating into five-branes.  Physically one
can imagine five-branes propagating between the two ends-of-the-world
in the M theory description, which corresponds to blowups, blowdowns
transforming ${\bf F}_{n}$ into ${\bf F}_{n+1}$.  (To avoid certain 
technical issues, we will assume $n \leq 6$.)

The transformation ${\bf F}_{n} \, \rightarrow \, {\bf F}_{n \pm 1}$ has
an elegant mathematical understanding as an elementary
transformation on the ruled surface ${\bf F}_{n}$ \cite{beauville}.  Viewed
as a ${\bf P}^{1}$ fibration, 
${\bf F}_{n}$ has 
sections, of self-intersection $\pm n$.  The transformation
proceeds by  
first blowing up a point on one of the sections, then blowing down the
strict transform of the fiber\footnote{Note if $S$ is the section
and $F$ the fiber, then the strict transform of the fiber $\hat{F} \, =
\, \pi^{*} F \, - \, E$ has self-intersection $-1$, as $E^{2} \, = \,
-1$, $E \cdot \pi^{*} F \, = \, 0$, and $F^{2} \, = \, 0$, 
so by Castelnuovo's contractibility criterion and the fact $\hat{F}
\cong {\bf P}^{1}$, one can blowdown $\hat{F}$ to recover a smooth
surface.}.

\begin{figure}
\centerline{\psfig{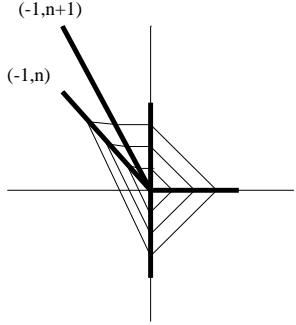}}
\caption{\label{oneblowup} A fan describing ${\bf F}_{n}$ with a
single blowup}
\end{figure}

To make this section accessible to a larger number of readers, we will use the fact that Hirzebruch surfaces and
some of their blowups are toric varieties\footnote{For introductions to
toric varieties see \cite{fulton,oda,russ,cox}.}.
The fan describing the surface relevant here as a toric variety,
a
(smooth toric) blowup of ${\bf F}_{n}$,
is shown in Figure \ref{oneblowup}.

The fan has edges along $(0,1)$, $(1,0)$, $(-1,n)$, $(-1,n+1)$, and $(0,-1)$.
Using standard methods, 
the toric variety can
be described as a GIT quotient.  If we also fiber ${\bf P}^{2}$
over the base, then we have homogeneous coordinates and ${\bf C}^{\times}$ actions
defining the total space as follows:

\begin{center}
\begin{tabular}{c|ccccc|ccc} 
 \, & $s$ & $t$ & $u$ & $v$ & $w$ & $x$ & $y$ & $z$ \\ \hline
$\lambda$ & 1 & 1 & $n$ & 0 & 0 & $2 \alpha$ & $3 \alpha$ & 0 \\
$\mu$ & 0 & 0 & 1 & 1 & 0 & $2 \beta$ & $3 \beta$ & 0 \\
$\tau$ & 1 & 0 & $n+1$ & 0 & 1 & $2 \gamma$ & $3 \gamma$ & 0 \\
$\nu$ & 0 & 0 & 0 & 0 & 0 & 1 & 1 & 1 
\end{tabular}
\end{center}
 
At this point a few words of explanation are in order.  The homogeneous
coordinates $x$, $y$, $z$ are coordinates on the fiber ${\bf P}^{2}$.
The coordinates $u$, $v$ can be understood as homogeneous coordinates
on the ${\bf P}^{1}$ fiber of the Hirzebruch surface
that exists in either of the blowdown limits to be
described next.  

To recover ${\bf F}_{n}$, we blowdown the toric divisor described by the
edge $(-1,n+1)$ in the fan, which corresponds to using the $\tau$ action
to fix the value of $w$ and taking $s$, $t$ to be homogeneous
coordinates on the ${\bf P}^{1}$ base of ${\bf F}_{n}$.  The exceptional
divisor is inserted at the point $v \, = \, t \, = \, 0$.

To recover ${\bf F}_{n+1}$, we blowdown the toric divisor described by the
edge $(-1,n)$ in the fan, which corresponds to using the $\lambda$
action to fix the value of $t$ and taking $s$, $w$ to be the homogeneous
coordinates on the ${\bf P}^{1}$ base of ${\bf F}_{n+1}$.

The Calabi-Yau three-fold is described as a hypersurface in the toric
four-fold above:
\begin{displaymath}
y^{2} z \: = \: x^{3} \: + \: f(s,t,u,v,w) x z^{2} \: + \: g(s,t,u,v,w)
z^{3}
\end{displaymath}
For this to describe a Calabi-Yau, we must demand $\alpha \, = \, n+2$,
$\beta \, = \, 2$, $\gamma \, = \, n+3$, and that $f$, $g$ have
degrees under the ${\bf C}^{\times}$ actions as below:
\begin{center}
\begin{tabular}{c|cc}
 \, & degree $f$ & degree $g$ \\ \hline
$\lambda$ & $4 \alpha$ & $ 6 \alpha$ \\
$\mu$ & $4 \beta$ & $6 \beta$ \\
$\tau$ & $4 \gamma$ & $6 \gamma$ 
\end{tabular}
\end{center}
so we can expand $f$, $g$ as
\begin{displaymath}
f(s,t,u,v,w) \: = \: \sum_{i=0}^{8} \, u^{i} v^{8-i} \, \left[ \, \sum_{j} \,
f_{i,j} \, s^{8+n(4-i)-j} \, t^{j} \, w^{4-i+j} \, \right]
\end{displaymath}
\begin{displaymath}
g(s,t,u,v,w) \: = \: \sum_{i=0}^{12} \, u^{i} v^{12-i} \, \left[ \,
\sum_{j} \, g_{i,j} \, s^{12 + n(6-i) - j} \, t^{j} \, w^{6-i+j} \,
\right]
\end{displaymath}

Now assume that we have a section of $E_{6}$ singularities at $v \, = \,
0$, which corresponds to an $SU(3)$ bundle in one of the $E_{8}$s.
Following \cite{btate,paulgross} this means that in a
neighborhood of $v \, = \, 0$, in terms of the affine coordinate $w_{1}
\, = \, v/u$, we have
\begin{displaymath}
f(w_{1},s,t,w) \: = \: w_{1}^{3} \, f'_{8-n}(s,t,w)
\end{displaymath}
\begin{displaymath}
g(w_{1},s,t,w) \: = \: w_{1}^{4} \, \left[ q'_{6-n}(s,t,w) \right]^{2}
\: + \: w_{1}^{5} \, g'_{12-n}(s,t,w)
\end{displaymath}
where
\begin{displaymath}
f'_{8-n}(s,t,w) \: = \: \sum_{j=1}^{8-n} \, f'_{8-n,j} \, s^{8-n-j} \,
t^{j} \, w^{j-1} 
\end{displaymath}
\begin{displaymath}
q'_{6-n}(s,t,w) \:= \: \sum_{j=1}^{6-n} \, q'_{6-n,j} \, s^{6-n-j} \,
t^{j} \, w^{j-1}
\end{displaymath}
\begin{displaymath}
g'_{12-n}(s,t,w) \: = \: \sum_{j=1}^{12-n} \, g'_{12-n,j} \, s^{12-n-j} \,
t^{j} \, w^{j-1}
\end{displaymath}

As explained in \cite{wmf}, in either blowdown limit, $f'$, $q'$, and $g'$ above define a
spectral curve over the ${\bf P}^{1}$ which is the base of the elliptic $K3$ (with
section) of the heterotic compactification.

Consider the blowdown limit to ${\bf F}_{n}$.  Were it not for the homogeneous
coordinate $w$ above, $f'$, $q'$, and $g'$ could each have an additional
term, with coefficients $f'_{8-n,0}$, $q'_{6-n,0}$, and $g'_{12-n,0}$.
However, in the blowdown limit $f'$, $q'$,
and $g'$ all have a common factor
of $t$, i.e., $f'$, $q'$, and $g'$ all vanish at $t \, = \, 0$.
In other words, in the blowdown limit in which we recover ${\bf F}_{n}$ as the
base of the elliptic fibration in F theory, the spectral curve
(partially) defining the vector bundle on $K3$ in the heterotic dual
becomes reducible, and now describes a sheaf,
not 
a vector bundle.  This vector
bundle degeneration coincides with the appearance of tensionless
strings in the compactified six-dimensional theory.

It is possible to streamline the derivation above.  Let
$\pi$ denote the blowdown morphism to ${\bf F}_{n}$, and $\pi': {\bf
F}_{n} \rightarrow {\bf P}^{1}$ the projection morphism, then if
$\tilde{{\cal L}}$ denotes the anticanonical bundle of the blowup of
${\bf F}_{n}$, $\pi_{*} \tilde{{\cal L}}^{4} \, = \, {\cal L}^{4}
\otimes {\cal I}_{4}$, where ${\cal I}$ is an ideal sheaf vanishing at
the point $t \, = \, v \, = \, 0$, and 
\begin{displaymath}
\pi'_{*} \pi_{*} \tilde{{\cal L}}^{4} \: = \: \bigoplus_{i = 0}^{3} {\cal
O}(8+n(4-i)) \, \bigoplus {\cal O}(8) \, \bigoplus_{i=5}^{8} \left[ {\cal O}(8+(n+1)(4-i)) \otimes {\cal
I}'_{i-4} \right]
\end{displaymath}
where ${\cal I}'$ is an ideal sheaf vanishing over $t = 0$.  (As a
check, note that $h^{0}({\bf P}^{1},{\cal O}(8)) \, = \, 9 \, = \,
\mbox{rank } \pi'_{*} \pi_{*} \tilde{{\cal L}}^{4}$.)  In this
fashion we can
read off the spectral data defining each bundle as well as the data
defining the heterotic Weierstrass fibration.

So far we have only studied the bundle moduli of the heterotic
compactification, in the limit that a tensionless string should appear
(by duality with F theory).  What about the moduli of the elliptic
fibration in the heterotic compactification, the $K3$ moduli?
The polynomials defining the heterotic Weierstrass fibration also appear
in the F theory Weierstrass polynomials:  if we expand
\begin{eqnarray*}
f(s,t,u,v,w) & = & \sum_{i=0}^{8} \, u^{i} v^{8-i} \, f_{i}(s,t,w) \\
g(s,t,u,v,w) & = & \sum_{i=0}^{12} \, u^{i} v^{12-i} \, g_{i}(s,t,w)
\end{eqnarray*}
then the polynomials appearing in the heterotic Weierstrass fibration
are precisely $f_{4}(s,t,w)$ and $g_{6}(s,t,w)$.  Now,
\begin{eqnarray*}
f_{4}(s,t,w) & = & \sum_{j} \, f_{4,j} \, t^{j} s^{8-j} w^{j} \\
g_{6}(s,t,w) & = & \sum_{j} \, g_{6,j} \, t^{j} s^{12-j} w^{j}
\end{eqnarray*}
so in the blowdown limit there are no additional constraints on these
polynomials.  In other words, in the limit in which a tensionless string
should appear, the bundle becomes singular, but the elliptic $K3$ need
not.  This is completely consistent with our understanding of the
physics, as we expect tensionless strings to appear when an instanton
shrinks to zero size -- $K3$ moduli should be irrelevant.

What is the dimension of the subvariety of $K3$ along which the bundle
degenerates?  Since the spectral data $a_{k}$ all vanish along a
codimension one locus, it is naively tempting to speculate that the
bundle must degenerate along some curve in the $K3$.  However,
there is an important subtlety -- it will turn out that the bundle does
not degenerate along an entire curve in $K3$, but at most at points.

What is this subtlety?  The fact that the $a_{k}$ all vanish along some
codimension one subvariety means that the $a_{k}$ are not really
sections of ${\cal N} \otimes {\cal L}^{-k}$, but rather sections
of ${\cal N} \otimes {\cal L}^{-k} \otimes {\cal M}$, for some line
bundle ${\cal M}$ (determined by the
divisor $\{ t \, = \, 0 \}$ in this case).  We can recover manifestly well-defined spectral data
simply by making some redefinitions: $a_{0} \: = \: g'_{12-n} \:
\rightarrow \: g'_{12-n} / t$, and so forth.  The new spectral data $a_{k}$
are sections of ${\cal N} \otimes {\cal L}^{-k}$.  In this particular
case, the spectral cover is reducible, with one component being the
spectral cover of a bundle with $c_{2} \, = \, 11-n \, = \, 12-n-1$,
reflecting the fact that a single instanton has become small.

To repeat, we have observed that if the spectral data $a_{k}$ vanishes
along some codimension one locus on the base of the elliptic fibration, it should not be interpreted as a
bundle degeneration along a codimension one locus, but rather as a 
poor interpretation of the spectral data.  Any naive codimension one
bundle degenerations should really be interpreted as, at most, 
codimension two. 
Note that the procedure outlined above only applies if the $a_{k}$
vanish at codimension one, not at codimension two.
 
The fact that the bundle does not degenerate at
codimension one on $K3$, but only at codimension two, is perfectly
consistent with the physical interpretation of the bundle degeneration as due to
the presence of a five-brane, localized at a point on $K3$.

The spectral curve degeneration occurs at complex codimension three in
the moduli space of spectral curves,
as we have lost one monomial in each of $f'$, $q'$, and $g'$.  What is the
codimension of the vector bundle degeneration (in the moduli space of
vector bundles on $K3$)?  Recall that the vector
bundle on $K3$ is defined by both a spectral cover of the base
of the elliptic $K3$ as well as a line bundle on the spectral cover.
In this case, since the base is ${\bf P}^1$, the spectral cover is a branched
cover of ${\bf P}^1$, i.e., a Riemann surface, and the space of line bundles
of fixed degree is of course the Jacobian of the Riemann surface.
If the Riemann surface degenerates at complex codimension three,
then surely the Jacobian will also
degenerate at codimension three\footnote{For every cycle in the Riemann
surface that shrinks, the space of flat line bundles loses precisely one
dimension.}, so the degenerate vector bundle moduli
space lies along a complex codimension six subvariety, which is
precisely correct to describe one less $SU(3)$ instanton on $K3$.

This degeneration of $f(s,t,u,v,w)$ and $g(s,t,u,v,w)$ also yields a
singular elliptic fiber (in the F theory compactification) over the point $t \, = \, 0$.  The singularity
is type $II$ in Kodaira's classification, which means that although
the fiber is singular, the total space is not singular in a neighborhood
of this point, and so there is no (nonperturbative) contribution to the six-dimensional
gauge symmetry, precisely as expected.

So far, we have discussed the degeneration of the $SU(3)$ bundle in the
$E_{8}$ located over $v \, = \, 0$.  Proceeding similarly, one can show
the bundle in the other $E_{8}$, located over $u \, = \, 0$, does
not degenerate.  This is completely consistent with the fact that we
have blown up ${\bf F}_{n}$ over the point $v \, = \, 0, \: t \, = \,
0$, so the bundle over the blown-up point degenerates, and the
spectral curve degenerates over the same point on the base.

We can also study the limit in Kahler moduli space in which we recover
${\bf F}_{n+1}$.  This limit corresponds to a blow-up of ${\bf F}_{n+1}$
over the point $u \, = \, 0, \: w \, = \, 0$.  The bundle over $u \, =
\, 0$ degenerates, and the spectral curve degenerates over the point
$w \, = \, 0$ on the base.  Note that all of this is completely
consistent with the standard interpretation of such blowups:  they
should correspond to an instanton in one $E_{8}$ shrinking and becoming
a five-brane, travelling to the second $E_{8}$, and reverting to an
instanton.

\subsection{Multiple Small Instantons}

\begin{figure}
\centerline{\psfig{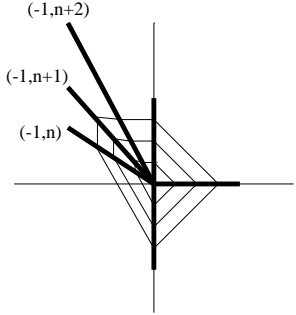}}
\caption{\label{twoblow} The fan describing ${\bf F}_{n}$ with two
blowups.} 
\end{figure}

In this subsection we will study spectral curve degeneration in a case 
in which two instantons shrink to
become five-branes, whose F-theory dual is an elliptic three-fold 
over ${\bf F}_{n}$ with two of blowups, as shown in Figure
\ref{twoblow}.  (Note that the second blowup is located on the
exceptional divisor of the first blowup.)  Closely related results have
been obtained in \cite{bs}.  The details proceed in a very similar manner to the last
section, so we will simply outline the relevant results.
 
If we fiber ${\bf P}^{2}$ over this base, then the complete
description in terms of homogeneous coordinates and ${\bf C}^{\times}$
actions is as follows:
\begin{center}
\begin{tabular}{c|cccccc|ccc}
\, & $s$ & $t$ & $u$ & $v$ & $w$ & $r$ & $x$ & $y$ & $z$ \\ \hline
$\lambda$ & 1 & 1 & $n$ & 0  & 0 & 0 & $2 \alpha$ & $3 \alpha$ & 0 \\
$\mu$ & 0 & 0 & 1 & 1 & 0 & 0 & $2 \beta$ & $3 \beta$ & 0 \\
$\tau$ & 1 & 0 & $n+1$ & 0 & 1 & 0 & $2 \gamma$ & $3 \gamma$ & 0 \\
$\rho$ & 1 & 0 & $n+2$ & 0 & 0 & 1 & $2 \delta$ & $3 \delta$ & 0 \\
$\nu$ & 0 & 0 & 0 & 0 & 0 & 0 & 1 & 1 & 1 \\
\end{tabular}
\end{center}

As before, we will describe our elliptic three-fold by the hypersurface
\begin{displaymath}
y^{2} z \: = \: x^{3} \: + \: f(s,t,u,v,w,r) \, x \, z^{2} \: + \:
g(s,t,u,v,w,r) \, z^{3}
\end{displaymath}
and for this to be a Calabi-Yau, we demand $\alpha \: = \: n+2$, $\beta
\: = \: 2$, $\gamma \: = \: n+3$, and $\delta \: = \: n+4$.

Suppose we have an $SU(3)$ bundle over $v \, = \, 0$, then the
Weierstrass fibration is of the form
\begin{displaymath}
f(w_{1},s,t,w,r) \: = \: w_{1}^{3} \, f'_{8-n}(s,t,w,r)
\end{displaymath}
\begin{displaymath}
g(w_{1},s,t,w,r) \: = \: w_{1}^{4} \, \left[ \, q'_{6-n}(s,t,w,r) \,
\right]^{2} \: + \: w_{1}^{5} \, g'_{12-n}(s,t,w,r)
\end{displaymath}
and we can expand each of the terms as
\begin{eqnarray*}
f'_{8-n}(s,t,w,r) & = & \sum_{i} \, f'_{8-n,i} \, s^{8-n-i} \, t^{i} \,
w^{i-1} \, r^{i-2} \\
q'_{6-n}(s,t,w,r) & = & \sum_{i} \, q'_{6-n,i} \, s^{6-n-i} \, t^{i} \,
w^{i-1} \, r^{i-2} \\
g'_{12-n}(s,t,w,r) & = & \sum_{i} \, g'_{12-n,i} \, s^{12-n-i} \, t^{i}
\, w^{i-1} \, r^{i-2}
\end{eqnarray*}

Thus, in the blowdown limit in which we recover ${\bf F}_{n}$, we find
that each of $f'$, $q'$, $g'$ is proportional to $t^{2}$, not just $t$.
A dimension count just like the one in the last section reveals that
this is perfect to describe the shrinking of two $SU(3)$ instantons on
$K3$.  

In the last section, when one instanton shrank,
there was no nonperturbative enhanced gauge symmetry -- the singularity
in the fiber was Kodaira type $II$.  Here, however, the singularity is
Kodaira type $IV$, which corresponds to an $A_{2}$ singularity in the
total space of the F-theory compactification, so we expect to recover
a nonperturbative enhanced $SU(3)$ gauge symmetry in six dimensions.

\section{Conclusions}

In this paper we developed technology to describe a class of extremal transitions 
directly in heterotic string
theory. 
We have also studied the sheaves
associated with small instantons on $K3$, and the corresponding spectral
cover degenerations in each case.

The work described here leaves many questions answered.  Perhaps
foremost among these questions concerns the nonperturbative physics at
the transitions, about which we have had very little to say.  Each of
the transitions between four-dimensional compactifications discussed in
this paper could conceivably be obstructed by a superpotential, an issue
we have not been able to address at all.  Without having a detailed
understanding of the physics occurring at these transitions, we have
only been able to make some preliminary checks of whether they might be
allowed -- by checking that the Dirac index is constant through the
transition, and that the number of five-branes present on both sides of
the transition is positive.  

However, in spite of not understanding the superpotential, we have found
that understanding classical heterotic extremal transitions is within reach of current
technology.

\section{Acknowledgements}

We would like to thank P. Aspinwall, A. Knutson, P. Mayr, D. Morrison, E. Witten
and especially T. Gomez 
for useful conversations.

\appendix

\section{Ideal Sheaves}

An ideal sheaf is simply a subsheaf of the trivial rank one sheaf (the
structure sheaf) with the property that all sections of the ideal sheaf
vanish along some subvariety.

More precisely, in a local coordinate neighborhood an ideal sheaf is
defined by some polynomial ideal, in the sense that all local sections
are elements of the ideal\footnote{Experts will recognize this is a
ham-handed treatment of a simple idea.  Let $(\mbox{Spec } A, {\cal O})$ be an
affine scheme, and $I$ an ideal of the ring $A$.  Then the stalk of the
ideal sheaf $\tilde{I}$ defined by $I$ over a prime ideal $p$ in $\mbox{Spec } A$ is the
localization $I_{p}$, and $\Gamma( \mbox{Spec } A, \tilde{I}) = I$.}.  For example, consider a local coordinate
neighborhood on some variety containing coordinates $u$, $v$, among
others.  An ideal sheaf that vanishes to first order at the point $u \,
= \, v \, = \, 0$ is defined by the ideal generated by $(u,v)$, i.e.,
local sections are all of the form $u f \, + \, v g$ for holomorphic
functions $f$, $g$.

In particular, an ideal sheaf that vanishes along a codimension one
locus is associated to a line bundle.

Chern classes can be defined for ideal sheaves.  To do so, one needs a
locally free resolution of the ideal sheaf.  For an ideal generated by a
regular sequence \cite{matsumura,hartshorne} (that is, a locally
complete intersection), we can use the Koszul resolution
\begin{displaymath}
\cdots \, \rightarrow \, \oplus {\cal O}(-D_{i} - D_{j}) \, \rightarrow
\, \oplus {\cal O}(- D_{i}) \, \rightarrow \, I \, \rightarrow 0
\end{displaymath}
where $I$ is the ideal sheaf, associated with the ideal generated by
the regular sequence $(f_{1}, f_{2}, \ldots)$, and $D_{i}$ are the divisors $D_{i} \, = \, \{
f_{i} = 0 \}$.  Then, given this resolution, we can compute the total
Chern class as
\begin{displaymath}
c(I) \: = \: c( \oplus {\cal O}(-D_{i}) ) \, c( \oplus {\cal O}(-D_{i}
-D_{j}))^{-1} \, \cdots
\end{displaymath} 

When an ideal sheaf is associated with an ideal not generated by a
regular sequence, one must work harder, as the naive Koszul resolution
does not yield an exact sequence.

Note that an ideal sheaf is not uniquely defined by the subvariety
along which it vanishes and the order to which it vanishes along this
subvariety.  To return to the previous example, consider the ideals
generated by $(u^{n},v^{n})$ and by $(u^{n},u^{n-1} v, u^{n-2}v^{2},
\ldots, v^{n})$.  Both of these vanish to order $n$ at the subvariety $u
\, = \, v \, = \, 0$.  However, the first is a regular sequence, the second is not, so
in general the ideal sheaves generated by either will have distinct
Chern classes.

\section{Homological Algebra}

In this appendix we will give an extremely schematic outline of some
homological algebra used in this paper.  For a basic introduction to the
subject, see \cite{hs,matsumura}, and for information on the
sheaf-theoretic version, see \cite{okonek,gh,hartshorne}.

Let $M$, $N$ be modules over some ring $A$, then a set of groups labelled
$\mbox{Ext}^{n}(M,N)$ can be
associated with the pair $(M,N)$, and classify isomorphism classes of
exact sequences
\begin{displaymath}
0 \: \rightarrow \: N \: \rightarrow \: E_{1} \: \rightarrow \: \cdots
\: \rightarrow \: E_{n} \: \rightarrow \: M \: \rightarrow \: 0
\end{displaymath}
Rather than explain the technical
definition, we will merely state some relevant properties:

1)  If the ring A is a principal ideal
domain, then $\mbox{Ext}^{n}(M,N) \, = \, 0$ for all $n$ and for all $N$
precisely when $M$ is freely generated.

2)  $\mbox{Ext}^{0}(M,N) \, = \, \mbox{Hom}(M,N)$

3)  If the following sequences are exact,
\begin{displaymath}
0 \: \rightarrow \: M' \: \rightarrow \: M \: \rightarrow \: M'' \:
\rightarrow \: 0
\end{displaymath}
\begin{displaymath}
0 \: \rightarrow \: N' \: \rightarrow \: N \: \rightarrow \: N'' \:
\rightarrow \: 0
\end{displaymath}
then we have the long exact sequences
\begin{displaymath}
\cdots \: \rightarrow \: \mbox{Ext}^{n}(M,N) \: \rightarrow \:
\mbox{Ext}^{n}(M',N) \: \rightarrow \: \mbox{Ext}^{n+1}(M'',N) \:
\rightarrow \: \cdots 
\end{displaymath}
\begin{displaymath}
\cdots \: \rightarrow \: \mbox{Ext}^{n}(M,N) \: \rightarrow \:
\mbox{Ext}^{n}(M,N'') \: \rightarrow \: \mbox{Ext}^{n+1}(M,N') \:
\rightarrow \: \cdots
\end{displaymath}
i.e., as a functor, $\mbox{Ext}^{n}(-,-)$ is contravariant in the first
variable and covariant in the second.

So far we have described $\mbox{Ext}$ for modules.  It is also possible
to define $\mbox{Ext}$ for sheaves, and in fact one recovers two
distinct possibilities, called local ${\it Ext}$ and global
$\mbox{Ext}$.

Local ${\it Ext}$ is a sheaf derived from a pair of coherent sheaves,
say ${\cal M}$ and ${\cal N}$.  It can be derived as $\mbox{Ext}$
for modules acting on stalks of the sheaves ${\cal M}$, ${\cal N}$,
and has the same properties as for modules: 

1)  If ${\cal M}$ is locally free, i.e., associated to a vector bundle,
then ${\it Ext}^{n}({\cal M},-) \, = \, 0$ for all $n > 0$.

2)  ${\it Ext}^{0}({\cal M},{\cal N}) \, = \, {\it Hom}({\cal M},{\cal
N})$

3)  And one gets long exact sequences of ${\it Ext}$ sheaves from short
exact sequences of sheaves as above.

In addition to the sheaf local ${\it Ext}$, it is also possible to
define a group, global $\mbox{Ext}$.  This group is defined as the limit
of either of two spectral sequences, with second level terms
\begin{displaymath}
E_{2}^{p,q} \: = \: H^{p}( {\it Ext}^{q}({\cal M},{\cal N}))
\end{displaymath}
\begin{displaymath}
E_{2}^{' p,q} \: = \: H^{q}( {\it Ext}^{p}({\cal M},{\cal N}))
\end{displaymath}

Isomorphism classes of exact sequences of sheaves
\begin{displaymath}
0 \: \rightarrow \: {\cal N} \: \rightarrow \: {\cal E} \: \rightarrow
\: {\cal M} \: \rightarrow \: 0
\end{displaymath}
are classified by elements of global $\mbox{Ext}^{1}({\cal M},{\cal
N})$.  In particular, if ${\cal M}$ is locally free, then
$\mbox{Ext}^{1}({\cal M},{\cal N}) \, = \, H^{1}({\it Hom}({\cal
M},{\cal N}))$, a result oft-mentioned in \cite{wmf}.

\end{document}